\documentclass[runningheads]{llncs}
\usepackage{amsmath}
\usepackage[ruled,vlined,linesnumbered]{algorithm2e}
\usepackage{amsmath}
\usepackage[T1]{fontenc}
\usepackage{graphicx}
\usepackage{textcomp}
\usepackage{tikz}
\usepackage{hyperref}
\usepackage{textcomp}
\usepackage{color}
\usepackage[utf8]{inputenc}
\usepackage{fancyvrb}
\usepackage{xcolor}
\usepackage{colortbl}
\usepackage{hhline}
\usepackage{lscape}
\usepackage{upquote}
\usepackage{array}
\usepackage{adjustbox}
\usepackage{rotating}
\usepackage{stmaryrd}
\usepackage{pifont}
\usepackage{multirow}
\usepackage{wrapfig}
\usepackage{verbatim}
\usepackage{colortbl}
\usepackage{rotating}

\newcolumntype{L}[1]{>{\raggedright\arraybackslash}m{#1}}
\newcolumntype{C}[1]{>{\centering\arraybackslash}m{#1}}
\newcolumntype{R}[1]{>{\raggedleft\arraybackslash}m{#1}}

\fvset{commandchars=\\\{\}}

\definecolor{Green}{rgb}{0.0, 0.56, 0.0}
\definecolor{Gray}{gray}{0.85}
\newcommand{\Blue}[1]{\textcolor{blue}{\textbf{#1}}}

\newcommand{\Red}[1]{\textcolor{red}{\textbf{#1}}}

\newcommand{\Var}[1]{\textit{#1}}

\newcommand{\CodeIn}[1]{\begin{small}\texttt{#1}\end{small}}
\newcommand{\Comment}[1]{}
\newcommand{\NoType}[1]{} 
\newcommand{\Space}[1]{}

\newcommand{\tool}{$Reach$}

\newcommand{\DefMacro}[2]{\expandafter\newcommand\csname rmk-#1\endcsname{#2}}
\newcommand{\UseMacro}[1]{\csname rmk-#1\endcsname}

\DefMacro{h1}{1H}
\DefMacro{h2}{2H}
\DefMacro{h3}{3H}
\DefMacro{h4}{4H}
\DefMacro{h5}{5H}
\DefMacro{h6}{6H}
\DefMacro{o1}{1OH}
\DefMacro{o2}{2OH}
\DefMacro{o3}{3OH}
\DefMacro{o4}{4OH}
\DefMacro{e1}{1EH}
\DefMacro{e2}{2EH}
\DefMacro{e3}{3EH}
\DefMacro{e4}{4EH}
\DefMacro{TwoTests}{2 Tests}
\DefMacro{FourTests}{4 Tests}
\DefMacro{EightTests}{8 Tests}
\DefMacro{SixteenTests}{16 Tests}
\DefMacro{EvalNone}{A$_{EB}$}
\DefMacro{EvalExpr}{A$_{EE}$}
\DefMacro{EvalSub}{A$_{ES}$}
\DefMacro{Solve}{A$_S$}
\DefMacro{TimeOut}{$\perp$}
\DefMacro{Error}{$\epsilon$}

\DefMacro{check}{\ding{51}}
\DefMacro{corr}{\ding{72}}
\DefMacro{plausible}{\ding{51}}

\DefMacro{NoPruning}{No Prun.}
\DefMacro{StaticPruning}{Static Prun.}
\DefMacro{DynamicPruning}{Dynamic Prun.}
\DefMacro{SmallTest}{Half Test Suite}
\DefMacro{LargeTest}{Full Test Suite}
\DefMacro{Unknown}{$\chi$}

\newcommand{\AC}[1]{// #1 \\}

\DefMacro{effectiveness}{Starting from Half Test Suite}
\DefMacro{inpractice}{To First Correct Solution}
\DefMacro{correct}{\# Correct}
\DefMacro{incorrect}{\# Incorrect}
\DefMacro{beforecorrect}{\# Incorrect}
\DefMacro{counterexamples}{\# Counterexamples}

\newcommand{\ACaret}{\raisebox{-0.6ex}{\textasciicircum}}
\newcommand{\AStar}{*}

\DefMacro{TotalModelNum}{12}
\DefMacro{model}{Model}
\DefMacro{problem}{Problem}
\DefMacro{expr-num}{\#expr}
\DefMacro{time}{time}
\DefMacro{ast-num}{\#AST}
\DefMacro{sig-num}{\#Sig}
\DefMacro{rel-num}{\#Rel}
\DefMacro{var-num}{\#Var}
\DefMacro{primary-var-num}{\#PVar}
\DefMacro{test-num}{\#Test}

\newcommand\diff\setminus

\urldef{\mailsa}\path|kaiyuanw@utexas.edu|
\urldef{\mailsb}\path|allisonksullivan@utexas.edu|
\urldef{\mailsc}\path|emmanouil.koukoutos@epfl.ch|
\urldef{\mailsd}\path|marinov@illinois.edu|
\urldef{\mailse}\path|khurshid@utexas.edu|

\usepackage{graphicx}
\begin{document}

\title{REACH: Refining Alloy Scenarios by Scope}

%
%
\author{Ana Jovanovic\inst{1} \and
  Allison Sullivan\inst{1} }

\authorrunning{A. Jovanovic, A. Sullivan}

\institute{The University of Texas at Arlington, USA\\
\email{ana.jovanovic@mavs.uta.edu, allison.sullivan@uta.edu}\\}
%
%

%
\maketitle              
\begin{abstract}
Writing declarative models has numerous benefits, ranging from automated reasoning and correction of design-level properties before systems are built, to automated testing and debugging of their implementations after they are built. Alloy is a declarative modeling language that is well suited for verifying system designs. A key strength of Alloy is its scenario-finding toolset, the Analyzer, which allows users to explore all valid scenarios that adhere to the model’s constraints up to a user-provided scope. In Alloy, it is common for users to desire to first validate smaller scenarios, then once confident, move onto validating larger scenarios. However, the Analyzer only presents scenarios in the order they are discovered by the SAT solver. This paper presents $Reach$, an extension to the Analyzer which allows users to explore scenarios by size. Experimental results reveal $Reach$'s enumeration improves performance while having the added benefit of maintaining a semi-sorted ordering of scenarios for the user. Moreover, we highlight $Reach$'s ability to improve the performance of Alloy's analysis when the user makes incremental changes to the scope of the enumeration. Reach is available at: \url{https://reachenumerator.github.io/} 
\end{abstract}
\section{Introduction}
Our lives are increasingly dependent on software systems. However, these same systems, even the most safety-critical ones, are notoriously buggy. Therefore, there is a growing need to produce reliable software while keeping the cost low. One solution is to make use of declarative modeling languages to help improve software correctness. Alloy~\cite{JacksonAlloy2002} is a first-order, relational logic that is used in industry and academia~\cite{CDDiff,CD2Alloy,OpenflowAlloy,Margrave,DynAlloy}.  A key strength of Alloy is the ability to develop models in the Analyzer, a scenario enumeration toolset that lets users explore behavior enabled by their models. To achieve this, the Analyzer translates Alloy models into KodKod~\cite{KodKod} formulas and invokes off-the-shelf Boolean satisfiability (SAT) solvers to search for scenarios, which are assignments to the sets of the model such that all executed formulas hold. As output, the Analyzer produces a collection of scenarios the user can iterate over. 

Scenario enumeration is a highly useful application of formal methods, which has been used to validate software designs \cite{Alluminum,Sullivan2017EvaluatingSM,AUnit,PorncharoenwaseETALCompoSat2018,JacksonAlloyBook2006}, provide model counting for reliability analysis of systems~\cite{FilieriETALICSE2013}, and to synthesis security attacks for
hardware architectures~\cite{CheckMateMICRO2018,CheckMateMicro2019,CheckMateGitHub}. However, effectively applying scenario enumeration relies heavily on the quality of scenarios generated and how easily users can explore the scenarios. Currently, the Analyzer's enumeration process is robust: the Analyzer will enumerate all scenarios up to the user provided scope. However, the order in which scenarios are discovered is based only on the order that the back-end SAT solver discovers them. Therefore, there is no guarantee that the next scenario enumerated will be in any way related to the current scenario. This random ordering  can discourage users from continuously interacting with the enumerator.
Additionally, when a user is trying to validate that their modeled constraints are correct, it is common for the user to desire to move from smaller to larger scenarios. Unfortunately, there is currently no native support for sorting scenarios by size within the Analyzer.  

To address this limitation, this paper introduces \tool, an extension to the Analyzer which provides support for staged generation of scenarios by size.  In this paper, we refer to the size of a scenario as the size of the largest signature in the scenario. Rather than producing one enumeration across the entire scope, \tool's execution results in a separate enumeration per size allowed by the scope. Therefore, \tool{} enables the user to intentionally explore scenarios of a specific size and to control when they move onto exploring scenarios of a larger size. To achieve this, 
\tool{} restricts the upper bound of the conjunction normal form (CNF) encoding produced by KodKod to match the size being enumerated and modifies the CNF encoding to enforce that a scenario meets the required size because of a specific signature in the model. 
As a result, \tool{} additionally applies a semi-sorted ordering to scenarios for the user based on the order signatures are declared in the model. We evaluate \tool{} in two ways. 
First, we explore the efficiency of our encoding in comparison to the standard Alloy enumeration and a baseline technique. These experiments reveal that our encoding has no noticeable overhead and can even improve the overall runtime. Second, we perform a case study in which we apply \tool{} in an iterative deepening setting, where the scope of the command is gradually increased, to highlight how \tool's functionality can  improve Alloy's analysis in this incremental setting.

In this paper, we make the following contributions:
\begin{itemize}
\item {\bf Staged Generation}.  We modify the Analyzer's execution environment to stage generation of scenarios by size. Correspondingly, we update the Analyzer's interfaces to support unique enumerations for each size.
\item {\bf Open Source}. We release our enumeration framework as an open-source extension of the latest stable release of the Analyzer. This allows users to gradually explore our new enumeration technique while still adhering to the original workflow of the Analyzer. \tool{} can be found at: \url{https://REACHEnumerator.github.io}.
\item {\bf Evaluation}.  We present an experimental evaluation using a variety of subject models used to evaluate recent advancements to Alloy. We evaluate the overhead of \tool's focuses search environment and compare \tool's performance to the default Analyzer enumeration and a baseline technique. 
\item {\bf Case Study}.  We explore how \tool{} can improve the performance for enumerating scenarios in an iterative deepening setting, a common model development practice.  

\end{itemize}
\section{Illustrative Example}\label{sec:example}
\subsection{Alloy}

To highlight how modeling in Alloy works, Figure~\ref{fig:alloy} depicts a model of a singly-linked list with an acyclic constraint. Signature paragraphs introduce atoms and their relationships (lines 1 - 2). Line 1 introduces a named set \CodeIn{List} and uses the relation \CodeIn{header} to express that each \CodeIn{List} atom has zero or one header nodes. Similarly, line 2 introduces the named set \CodeIn{Node} and uses the \CodeIn{link} relation to express that each \CodeIn{Node} atom points to zero or one other nodes.  Predicate paragraphs introduce named first order logic formulas that can be invoked elsewhere (lines 3 - 5). The predicate \CodeIn{Acyclic} uses universal quantification (\CodeIn{all}), set exclusion (\CodeIn{!in}), transitive closure (\CodeIn{\ACaret}) and reflexive transitive closure (\CodeIn{\AStar}) to express the idea that ``for every list, all nodes in the list are not reachable from themselves following one or more traversal down their \CodeIn{link} relation.'' Commands indicate which formulas to invoke and what scope, bound on the universe of discourse, to explore (line 6). The scope places an upper bound on the size of all signature sets. The command on line 6 asks the \CodeIn{Analyzer} to search for satisfying assignments to all the sets of the model (\CodeIn{List}, \CodeIn{header}, \CodeIn{Node}, and \CodeIn{link}) such that the \CodeIn{Acyclic} predicate is true using up to 3 \CodeIn{List} atoms and 3 \CodeIn{Node} atoms.

\usetikzlibrary{shapes.geometric, arrows}
\tikzstyle{arrow} = [line width=1.5pt,->,>=stealth]
\tikzstyle{darrow} = [line width=1.5pt,<->,>=stealth]
\tikzstyle{NodeAtom} = [rectangle, minimum width=.75cm, minimum height=.75cm, text centered, draw=black, fill=yellow!75!orange]
\tikzstyle{ListAtom} = [rectangle, minimum width=.75cm, minimum height=.75cm, text centered, draw=black, fill=orange!75!yellow]

\begin{figure}[!t]
    \centering

\begin{Verbatim}[]
1. \Blue{sig} List \{header: \Blue{lone} Node\}
2. \Blue{sig} Node \{link: \Blue{lone} Node\} 
3. \Blue{pred} acyclic \{
4.   \Blue{all} l : List | \Blue{all} n : l.header.*link | n !\Blue{in} n.^link 
5. \}
6. \Blue{run} \{acyclic\} \Blue{for} \Red{3}
\end{Verbatim}
\vspace{-4ex}
    \caption{Alloy Model of a Singly-Linked List}
    \label{fig:alloy}
    \vspace{1ex}
\end{figure}

Figure~\ref{fig:scenarios} (a), (b) and (c) graphically displays the first three scenarios found by the Analyzer when the command at line 6 is executed: an empty list with a disconnected node that has a cycle, a list with one node and a disconnected node that has a cycle, and two lists each with one node and no cycles.  These scenarios are iteratively discovered by the back-end SAT solver. 
In Alloy, scenarios are produced by KodKod, which generates a DIMACS formatted CNF formula equivalent to the invoked constraints and uses a back-end SAT solver to find a satisfying solution. KodKod then translates this solution back into an Alloy scenario.

Specifically, KodKod creates a list of all possible atom assignments that can populate a set and ties those to unique integer values, which make up the primary variables of the CNF formula passed to the SAT solver. While solving the CNF formula, the SAT solver determines if each primary variable should be positive (true) or negative (false). A positive assignment means the atom is in the scenario. To illustrate, for the command in Figure~\ref{fig:alloy}, KodKod produces 24 primary variables for the CNF encoding: variables 1-3 relate to set \CodeIn{List}, 4-6 relate to set \CodeIn{Node}, 7-15 relate to  set \CodeIn{header} and 16-24 relate to set \CodeIn{link}. For set \CodeIn{List}, the primary variable ``1'' reasons over whether or not atom \CodeIn{L0} is in set \CodeIn{List}, primary variable ``2'' reasons over whether or not atom \CodeIn{L1} is in set \CodeIn{List}, and primary variable ``3'' reasons over whether or not atom \CodeIn{L2} is in set \CodeIn{List}. If the SAT solver finds a solution in which ``1'' is positive, then \CodeIn{L0} will be in set \CodeIn{List} for the scenario. 
To enumerate a new scenario, KodKod instructs the SAT solver to find another satisfying assignment to the primary variables with an additional CNF clause that asserts that any new assignment found must differ from all previously discovered assignments by at least one primary variable.

\usetikzlibrary{shapes.geometric, arrows}
\tikzstyle{arrow} = [line width=1.5pt,->,>=stealth]
\tikzstyle{darrow} = [line width=1.5pt,<->,>=stealth]
\tikzstyle{NodeAtom} = [rectangle, minimum width=.75cm, minimum height=.75cm, text centered, draw=black, fill=yellow!75!orange]
\tikzstyle{ListAtom} = [rectangle, minimum width=.75cm, minimum height=.75cm, text centered, draw=black, fill=orange!75!yellow]

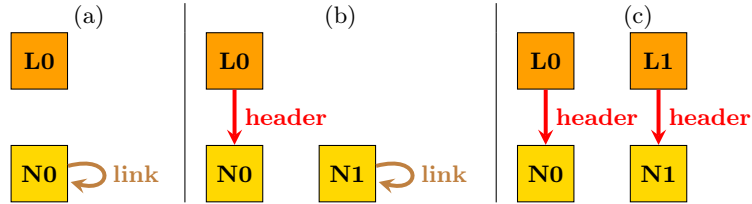
\begin{figure}[!t]
    \centering

\begin{tabular}[!t]{c|c|c}
\footnotesize (a) & \footnotesize (b) & \footnotesize (c) \\

\begin{minipage}[t]{.2\columnwidth}
\begin{center}
\begin{tikzpicture}[baseline,node distance=1.5cm]
\node (L0) [ListAtom] {\textbf{L0}};
\node (N0) [NodeAtom, below of=L0] {\textbf{N0}};
\begin{scope}[brown]
\draw [arrow] (N0) edge [loop right,line width=1.5pt] node {\textbf{link}} (N0);
\end{scope}[brown]
\end{tikzpicture}
\end{center}
\end{minipage}

&

\begin{minipage}[t]{.33\columnwidth}
\begin{center}
\begin{tikzpicture}[baseline,node distance=1.5cm]
\node (L0) [ListAtom] {\textbf{L0}};
\node (N0) [NodeAtom, below of=L0] {\textbf{N0}};
\node (N1) [NodeAtom, right of=N0] {\textbf{N1}};
\begin{scope}[red]
\draw [arrow] (L0) -- node[anchor=west] {\textbf{header}} (N0);
\end{scope}[red]
\begin{scope}[brown]
\draw [arrow] (N1) edge [loop right,line width=1.5pt] node {\textbf{link}} (N1);
\end{scope}[brown]
\end{tikzpicture}
\end{center}
\end{minipage}

&

\begin{minipage}[t]{.3\columnwidth}
\begin{center}
\begin{tikzpicture}[baseline,node distance=1.5cm]
\node (L0) [ListAtom] {\textbf{L0}};
\node (L1) [ListAtom, right of=L0] {\textbf{L1}};
\node (N0) [NodeAtom, below of=L0] {\textbf{N0}};
\node (N1) [NodeAtom, right of=N0] {\textbf{N1}};
\begin{scope}[red]
\draw [arrow] (L0) -- node[anchor=west] {\textbf{header}} (N0);
\draw [arrow] (L1) -- node[anchor=west] {\textbf{header}} (N1);
\end{scope}[red]
\begin{scope}[brown]

\end{scope}[brown]
\end{tikzpicture}
\end{center}
\end{minipage}
\end{tabular}
    \caption{Alloy Scenarios of a Singly-Linked List}
    \label{fig:scenarios}
\end{figure}

\subsection{\tool{} Enumeration}
The Analyzer's default enumeration process prevents repeatedly discovering duplicate scenarios; however, it does not guarantee any sort of ordering to the scenarios.
Therefore, \tool{} modifies Alloy's enumeration process to enumerate scenarios by size up to the user provided scope. In \tool, the user has the freedom to explore scenarios of any size and can choose when to switch between enumerating scenarios of different sizes.
To illustrate \tool's output, Figure~\ref{fig:reach} shows the first 8 scenarios enumerated by \tool{} if the user enumerates from the smallest size to the largest size.  To start, the user can iterative over all scenarios of size 0, which is just Figure~\ref{fig:reach}~(a). Then, the user can elect to iterate over all scenarios of size 1, which are the scenarios shown in Figure~\ref{fig:reach} (b) - (g). Next, the user can move onto scenarios of size 2, starting with the scenario in Figure~\ref{fig:reach} (h). For the model in Figure~\ref{fig:alloy}, \tool{} enumerates 1 scenarios of size 0, 6 scenarios of size 1, 38 scenarios of size 2 and 299 scenarios of size 3, with an average scenario discovery time of 2 milliseconds and a total runtime of 702 milliseconds. 

Besides sorting scenarios by size, \tool{} also creates a semi-sorted ordering of scenarios by enumerating scenarios based on which signature establishes the size of the scenario. To illustrate, the scenarios in Figure~\ref{fig:reach}~(b) - (e) are enumerated first, since \tool{} first explores all scenarios that are of size 1 because of the \CodeIn{List} signature. Then, once all those scenarios are enumerated, \tool{} explores all scenarios of size 1 becuase of the \CodeIn{Node} signature (Figure~\ref{fig:reach}~(f) - (g)).  In addition to appending CNF constraints which enforce the size of specific signatures, \tool{} still appends the CNF clause that prevents discovering duplicate scenarios. As a result, when exploring scenarios related to the signature \CodeIn{Node}, \tool{} does not rediscover scenarios Figure~\ref{fig:reach}~(a), (d), and (e). 
\section{Technique}
In this section, we describe \tool's CNF encoding (Section~\ref{sec:encoding}) and how we apply this encoding to staged generation of scenarios by size (Section~\ref{sec:design}). 

\usetikzlibrary{shapes.geometric, arrows}
\tikzstyle{arrow} = [line width=1.5pt,->,>=stealth]
\tikzstyle{darrow} = [line width=1.5pt,<->,>=stealth]
\tikzstyle{NodeAtom} = [rectangle, minimum width=.75cm, minimum height=.75cm, text centered, draw=black, fill=yellow!75!orange]
\tikzstyle{ListAtom} = [rectangle, minimum width=.75cm, minimum height=.75cm, text centered, draw=black, fill=orange!75!yellow]

\begin{figure}[!t]
    \centering
    
\begin{tabular}[t]{c|c|c|c}
\footnotesize (a) & \footnotesize (b) & \footnotesize (c) & \footnotesize (d)\\

\begin{minipage}[t]{.1\columnwidth}
\begin{center}
\vspace{1mm}
\Large
$\emptyset$ 
\end{center}
\end{minipage}

&
\begin{minipage}[t]{.2\columnwidth}
\begin{center}
\begin{tikzpicture}[baseline,node distance=1.5cm]
\node (L0) [ListAtom] {\textbf{L0}};
\node (N0) [NodeAtom, below of=L0] {\textbf{N0}};
\begin{scope}[red]
\end{scope}[red]
\begin{scope}[brown]
\draw [arrow] (N0) edge [loop right,line width=1.5pt] node {\textbf{link}} (N0);
\end{scope}[brown]
\end{tikzpicture}
\end{center}
\end{minipage}

&

\begin{minipage}[t]{.15\columnwidth}
\begin{center}
\begin{tikzpicture}[baseline,node distance=1.5cm]
\node (L0) [ListAtom] {\textbf{L0}};
\begin{scope}[red]
\end{scope}[red]
\end{tikzpicture}
\end{center}
\end{minipage}

&

\begin{minipage}[t]{.25\columnwidth}
\begin{center}
\begin{tikzpicture}[baseline,node distance=1.5cm]
\node (L0) [ListAtom] {\textbf{L0}};
\node (N0) [NodeAtom, below of=L0] {\textbf{N0}};
\begin{scope}[red]
\end{scope}[red]
\begin{scope}[brown]
\end{scope}[brown]
\end{tikzpicture}
\end{center}
\vspace{1mm}
\end{minipage}

\\ \hline

\footnotesize (e) & \footnotesize (f) & \footnotesize (g) & \footnotesize (h)\\

\begin{minipage}[t]{.15\columnwidth}
\begin{center}
\begin{tikzpicture}[baseline,node distance=1.5cm]
\node (L0) [ListAtom] {\textbf{L0}};
\node (N0) [NodeAtom, below of=L0] {\textbf{N0}};
\begin{scope}[red]
\draw [arrow] (L0) -- node[anchor=west] {\textbf{header}} (N0);
\end{scope}[red]
\begin{scope}[brown]
\end{scope}[brown]
\end{tikzpicture}
\end{center}
\end{minipage}

& 

\begin{minipage}[t]{.2\columnwidth}
\begin{center}
\begin{tikzpicture}[baseline,node distance=1.5cm]
\node (N0) [NodeAtom, below of=L0] {\textbf{N0}};
\begin{scope}[red]
\end{scope}[red]
\begin{scope}[brown]
\draw [arrow] (N0) edge [loop right,line width=1.5pt] node {\textbf{link}} (N0);
\end{scope}[brown]
\end{tikzpicture}
\end{center}

\end{minipage}

&

\begin{minipage}[t]{.15\columnwidth}
\begin{center}
\begin{tikzpicture}[baseline,node distance=1.5cm]
\node (N0) [NodeAtom, below of=L0] {\textbf{N0}};
\begin{scope}[red]
\end{scope}[red]
\begin{scope}[brown]
\end{scope}[brown]
\end{tikzpicture}
\end{center}
\end{minipage}

&

\begin{minipage}[t]{.27\columnwidth}
\begin{center}
\begin{tikzpicture}[baseline,node distance=1.5cm]
\node (L0) [ListAtom] {\textbf{L0}};
\node (N0) [NodeAtom, below of=L0] {\textbf{N0}};
\node (L1) [ListAtom, right of=L0] {\textbf{L1}};
\node (N1) [NodeAtom, below of=L1] {\textbf{N1}};
\begin{scope}[red]
\draw [arrow] (L0) -- node[anchor=west] {\textbf{header}} (N0);
\draw [arrow] (L1) -- node[anchor=west] {\textbf{header}} (N1);
\end{scope}[red]
\begin{scope}[brown]
\end{scope}[brown]
\end{tikzpicture}
\end{center}
\end{minipage}

\end{tabular}
\caption{Reach Enumeration}
    \label{fig:reach}
\end{figure}
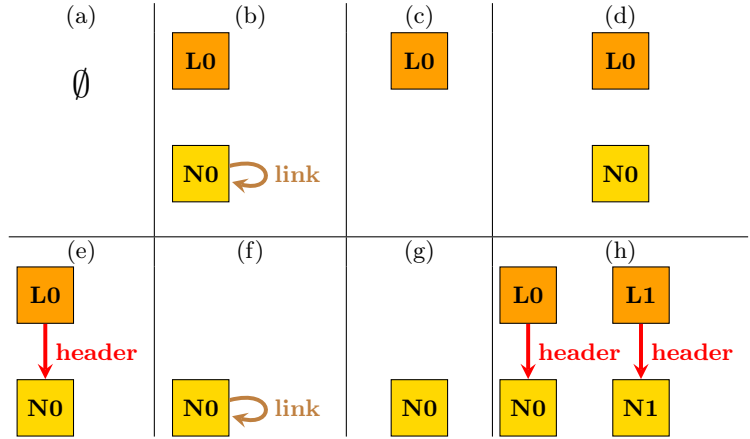

\subsection{CNF Encoding}\label{sec:encoding}
\tool{} creates new CNF clauses that enforce the size of a \emph{specific} signature. There are two main benefits of this. First, our CNF encoding is simplified, as we don't need to build a series of clauses to express the idea ``size of sig a is equal to Z or size of sig b is equal to Z or .... size of sig x is equal to Z,'' which is in a disjunctive normal form that is more complicated to translate into a equivalent CNF formula. Second, this allows us to automatically present scenarios in a semi-ordered format, as shown in Figure~\ref{fig:reach} (b) - (g), which can help users more confidently navigate the scenario space. To build our encoding, \tool{} appends a sequence of CNF clauses that asserts each primary variable mapped to the targeted signature is positive, meaning that the atom represented by the primary variable must appear in the signature for any scenario enumerated. To illustrate, to enumerate scenarios of size 3 due to the signature \CodeIn{List} using the primary variables outlined in Section~\ref{sec:example}, \tool{} would append the following clauses: $\{1\}$, $\{2\}$ and $\{3\}$. These clauses will ensure that atoms \CodeIn{L0}, \CodeIn{L1} and \CodeIn{L2} are present in set \CodeIn{List} for any scenario enumerated.

As an optimization, since \tool{} creates a distinct enumeration for each size within scope, \tool{} restricts KodKod's CNF generation to place an upper bound on all signatures that matches the target size being enumerated. Therefore, \tool's CNF encoding does not need to append any constraints for the other signatures present in the model, as KodKod's translation will ensure that the size of all other signatures falls between zero and the targeted size, inclusive. As an example, when enumerating scenarios of size 1 for the command in Figure~\ref{fig:alloy}, \tool{} tells KodKod to generate its CNF translation of the model with an upper bound of 1. As a result, KodKod will only create one primary variable for each signature. In contrast, the default translation, outlined in Section~\ref{sec:example}, would produce 3 primary variables for each signature. As a result, to enumerate only size 1 scenarios, \tool{} would need to append additional CNF clauses to prevent the size of all signatures from being too large. 



\subsection {Staged Generation}\label{sec:design}
To stage generation by size and to enforce an ordering on signatures, \tool{} modifies the Analyzer's  enumeration algorithm in two ways. First, as mentioned, \tool{} creates a separate enumeration \emph{per} size allowed by the scope. Therefore, \tool{} invokes KodKod multiple times, once for each size, to translate the model into a CNF formula. 
Second, when searching for scenarios, \tool{} adds the CNF clauses from Section~\ref{sec:encoding} that explicitly enforce the size a given signature. This results in an iterative process where \tool{} may need to update the CNF clauses when transitioning from focusing on one signature to the next. Currently, \tool{} prioritizes signatures based on the order in which they are declared in the model.  

\begin{algorithm}[!t]
  \small
  \caption{Staged Generation by Signature \label{fig:algo}}
  \DontPrintSemicolon
\SetKw{Continue}{continue}
\SetKw{Break}{break}
\SetKw{true}{True}
\SetKw{false}{False}
\SetKw{null}{null}
\KwIn{Kodkod solver \Var{solver}, Previous scenarios \Var{prev}, Set of signatures \Var{sigs},}
\KwOut{Set of scenarios of the target size.}

activeSig = 0, addClauses = false \AC{Initialize}
\While{activeSig != sigs.size()}{
    \If{addClauses}{
    \AC{Enact Reach's CNF encoding over relevant primary variables}
          \For{$i\gets1$ \KwTo sigs.get(activeSig).getPVariables().size()}{
                solver.addClause(sigs.get(activeSig).getPVariables().get(i))
          }
        addClauses = false
    }
    \Var{solution} = \Var{solver}.solve() ~~ \AC{Find next scenario} 
    \If{\Var{solution} != \null} {
        prev.add(solution.getPreventDuplicateClause())\\
        cnf.addClause(solution.getPreventDuplicateClause())\\
        output(\Var{solution})
    }
    \If{\Var{solution} == \null} {
        activeSig++\\ 
        solver.cnfClear()  ~~ \AC{Clear to remove previous active sig clauses}
        \For{$i\gets1$ \KwTo \Var{prev}.size() }{
            \Var{solver}.addClause(\Var{prev}.get(i)) ~~ \AC{Add clauses to prevent dups} 
        }
        addClauses = true\\
    }
} 
\end{algorithm}

Our modified enumeration approach is outlined in Algorithm~\ref{fig:algo}. The variable \CodeIn{activeSig} is used to keep track of which signature is the current focus of the enumeration. The boolean flag \CodeIn{addClauses} is used to determine if \tool{} should append new CNF clauses that assert the size of the signature indicated by \CodeIn{activeSig}, following the encoding outlined in Section~\ref{sec:encoding} (lines 4-7). Then, to search for the next scenario, \tool{} invokes the SAT solver on line 8. If the result is satisfiable, \tool{} generates the CNF clause to prevent rediscovering this scenario and outputs the scenario (lines 9 - 12). Once the SAT solver returns an unsatisfiable result (line 13), \tool{} updates \CodeIn{activeSig} to move onto the next signature, removes all CNF clauses and adds back the CNF clauses which prevent duplicate scenarios from being found (lines 16 - 17). Then, \tool{} sets the flag \CodeIn{addClauses} to \CodeIn{true}, which will trigger \tool{} to add size-based clauses for the next signature now indicated by \CodeIn{activeSig}. Once all signatures have been explored fully, the enumeration ends. 


To illustrate, to enumerate scenarios of size 1 for the model in Figure~\ref{fig:alloy}, \tool{} will translate the command into KodKod with an upper bound of 1, which means that KodKod will generate just 4 primary variables instead of 24: ``1'' reasons over if \CodeIn{L0} is in \CodeIn{List}, ``2'' reasons over if \CodeIn{N0} is in \CodeIn{Node}, ``3'' reasons over if \CodeIn{N0} is in \CodeIn{L0}'s \CodeIn{header} relation and ``4'' reasons over if \CodeIn{N0} is in \CodeIn{N0}'s \CodeIn{link} relation. Then, \tool{} will append the following clause, which assert that the size of \CodeIn{List} must be 1: $\{1\}$. This will result in the scenarios (b) - (e) in Figure~\ref{fig:reach}.
Once the SAT solver is no longer able to find any scenarios, \tool{} removes the clause $\{1\}$ and append the clause $\{2\}$, which assert that \CodeIn{N0} must be in \CodeIn{Node}, resulting in the scenarios in Figure~\ref{fig:reach} (f) and (g). Once the next iteration returns an unsatisfiable result, the enumeration ends and the user is informed that there are no more scenarios of size 1. Then, should the user wishes to explore scenarios of size 2, \tool{} would initiate a separate enumeration. This enumeration would be reasoning over the 12 primary variables generated by KodKod's translation. \tool{} would start by appending the following clauses, $\{1\}$ and $\{2\}$, which would ensure that atoms \CodeIn{L0} and \CodeIn{L1} are in the set \CodeIn{List}, respectively (e.g. Figure~\ref{fig:reach} (h)). 


\section{Implementation}
\tool{} is implemented as an extension to the Analyzer v.5.0.1~\cite{AlloyWebsite}.
Importantly, since \tool{} extends the main IDE for Alloy, users can directly explore the new functionality \tool{} provides within their normal development process. 

\subsection{Updating Alloy's Workflow}
\tool{} augments the enumeration workflow in the Analyzer, as seen in red in Figure~\ref{fig:framework}, which is the result of two main changes. First, \tool{} updates the communication between the Analyzer and KodKod, which is done through the Analyzer's Reporter, to include information about the target size. This enables KodKod to change the upper bound for its CNF translation to match that of the size actively being enumerated. Additionally, the Enumerator's iterator is updated to support the staged generation technique outlined in Algorithm~\ref{fig:algo}. 

\begin{figure}[!t]
    \centering
    \usetikzlibrary{positioning,fit,arrows.meta,backgrounds,shapes.geometric}

\tikzset{
    module/.style={%
        draw, rounded corners,
        minimum width=20mm,
        minimum height=7mm,
        },
    module/.default=2cm,
    >=LaTeX
}

\tikzset{
    scene/.style={%
        draw, 
        minimum width=12mm,
        minimum height=8mm,
        },
    module/.default=2cm,
    >=LaTeX
}

\tikzset{
    label/.style={%
        minimum width=#1,
        minimum height=5mm,
        },
    module/.default=1cm,
    >=LaTeX
}

\tikzset{datashape/.style={
  trapezium, draw, trapezium left angle=60,
  trapezium right angle=-60, minimum height=8mm}
  }

\tikzstyle{arrow} = [line width=1pt,->,>=stealth]
\tikzstyle{darrow} = [line width=1pt,<->,>=stealth]
\tikzstyle{darrow2} = [line width=0pt,<->,>=stealth]
\footnotesize
\begin{tikzpicture}[baseline,node distance=1.5cm,show background rectangle]

    \node[module] (I1) {Translator};
    \begin{scope}[red]
    \node[module, below=6mm of I1] (I2) {Reporter};
    \end{scope}[red]
    \node[module, below=6mm of I2] (I3) {VizGUI};
    \node[label, below=1mm of I3] (I4) {Analyzer};
    \node[fit=(I1) (I2) (I3) (I4), draw, inner sep=2mm] (fit1) {};
    
    \node[module, right=2cm of {I1-|fit1.east}] (I5) {Translator};
        \begin{scope}[red]
    \node[module, below=6mm of I5] (I6) {Enumerator};
    \end{scope}[red]
    \node[label, below=1mm of I6] (I7) {KodKod};
    
    \node[fit=(I5) (I6) (I7), draw, inner sep=2mm] (fit2) {};
    
    \node[module, below=8mm of I7] (I8) {SAT Solver};
    
    \node[scene, left=1.5cm of I1] (I10) {Model};
\node[scene,  left=1.5cm of I3] (I11) {Scenario};
    
    \begin{scope}[red]
\draw [arrow] (I1) -- node[anchor=west] {size} (I2);
\draw [darrow] (I2) -- node[anchor=west] {size} (I3);
\draw [darrow] (I5) -- node[anchor=west] {size} (I6);
\draw[arrow] ([yshift=0.8 cm]fit1)--  node[anchor=south] {size} (fit2);
\end{scope}[red]
\draw[arrow] ([yshift=-0.8 cm]fit2)--  node[anchor=north] {scenario} (fit1);

\draw [arrow] (fit1) -- node[anchor=west] {} (I11);
\draw [arrow] ([yshift=1 cm]I10) -- node[anchor=west] {} (fit1);
    \begin{scope}[red]
\draw [arrow] (fit2) -- node[anchor=west] {cnf} (I8);
\end{scope}[red]
\end{tikzpicture}
    \caption{\tool{} Framework Overview}
    \label{fig:framework}
\end{figure}

\begin{figure}[!t]
    \centering
    \includegraphics[scale = 0.5]{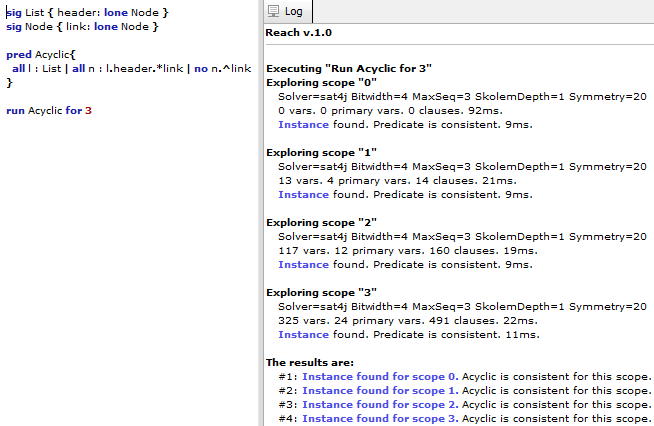}
        \caption{\tool{} Interface}
    \label{fig:interface}
\end{figure}

Second, in the Analyzer, a user is only allowed to enumerate \emph{one} command at a time. For \tool, we support multiple simultaneous enumerations, which can be thought of as enumerating multiple commands. Therefore, \tool{} modifies the interaction between the Reporter and the VizGUI interface to enable more than one concurrent enumeration. To support this, \tool{} updates the Analyzer's logging panel to display information per size, as shown in Figure~\ref{fig:interface}. When the user clicks on any enumeration link, the Analyzer's standard VizGUI interface appears, with the only modification being the functionality of the backend Enumerator. A user is able to switch between different enumerations by opening the respective VizGUI and a user can have multiple enumerations active at once. As a result, \tool{} improves the functionality of the original Analyzer. Specifically, the original Analyzer supports an ``execute all commands'' functionality; however, the Analyzer only allows the user to enumeration that last solved command. Our support for multiple simultaneous enumerations enables the user to be able to enumerate any of the solved commands.

\subsection{Supporting Alloy's Diverse Signature Grammar}
Alloy's robust grammar allows for signatures to be abstract, extensions of other signatures and to have their own multiplicity constraints. Some of these constructs do not result in primary variables for the SAT solver. In particular, if a signature has the singleton multiplicity constraint  (``\CodeIn{one}''), the constraint always results in the same atom assignment to the signature's set. Therefore, no primary variables are generated for this signature as there is nothing for the SAT solver to try and solve about what atoms should populate this signature. Unfortunately, this encoding behavior results in scenarios found for a scope that may the defy user's intention. To illustrate, if you have a signature with a singleton multiplicity constraint in your model and you execute a command with a scope of 0, the Analyzer will still find scenarios even though the signature with the singleton constraint will have a size of 1. To handle this, \tool{} appends a fact to the model which asserts the size of the signature is one, which prevents a scenario from being found of size 0. Therefore, \tool{} will  discover these scenarios as scenarios of size of 1, which we believe better matches the user's expectation. 

Abstract signatures also do not generate primary variables, as an abstract signature can never directly have atoms of its own. Instead, all atoms of any signature that extends the abstract signature count towards the size of the abstract signature. Unfortunately, an abstract signature can be extended by more than one signature, which does not work with \tool's streamlined CNF encoding because we have to account for all the different combinations of atoms from all extension signatures. Therefore, \tool{} instead handles this situation at the Alloy level. When enumerating scenarios that are of the target size for an abstract signature, \tool{} appends a fact that asserts that the size of the abstract signature matches the target size and lets KodKod generate the CNF encoding. 

\section{Evaluation}\label{sec:eval}
We evaluate \tool{} over a benchmark of Alloy models collected from recent advancements made to Alloy~\cite{Amalgam,ARepair,beafix}. We address the following research questions in this section:
\begin{itemize} \itemsep=0em
    \item \textbf{RQ1:} What is the overhead of \tool's signature-focused enumeration?
    \item \textbf{RQ2:} How does \tool's enumeration perform in comparison to the Analyzer's default enumeration and a baseline technique?
\end{itemize}

\subsection{Set Up}
\begin{wraptable}{r}{6cm}
\vspace{-5ex}
  \scriptsize
  \centering
  \caption{Size of evaluation models.}
  \label{tab:size}
  \begin{tabular}{lrrrrr}
\hline
\textbf{Model} & \textbf{\#Sig} & \textbf{\#Abs} & \textbf{\#Ext} & \textbf{\#One} & \textbf{\#Rel} \\ \hline
\rowcolor{Gray}
arr & 2 & 0 & 0 & 1 & 2 \\ 
\rowcolor{white}
btree & 2 & 0 & 0 & 1 & 4 \\ 
\rowcolor{Gray}
cd & 2 & 0 & 1 & 1 & 1 \\ 
\rowcolor{white}
CD2Alloy1 & 16 & 4 & 12 & 8 & 1 \\ 
\rowcolor{Gray}
CD2Alloy2 & 10 & 4 & 6 & 2 & 1 \\ 
\rowcolor{white}
diff1 & 17 & 4 & 13 & 8 & 1 \\ 
\rowcolor{Gray}
diff2 & 17 & 4 & 13 & 8 & 1 \\ 
\rowcolor{white}
dll & 2 & 0 & 0 & 1 & 4 \\ 
\rowcolor{Gray}
frankenvrep & 52 & 17 & 35 & 28 & 16 \\ 
\rowcolor{white}
nqueens & 1 & 0 & 0 & 0 & 2 \\ 
\rowcolor{Gray}
sll & 2 & 0 & 0 & 0 & 2 \\ 
\rowcolor{white}
tombot & 52 & 17 & 35 & 28 & 16 \\ \hline
\end{tabular}
  \vspace{-3ex}
\end{wraptable}

To determine the runtime impact of \tool, we enumerate scenarios across different models using (1)~the Analyzer, (2)~a baseline implementation of \tool's functionality and (3)~\tool. To collect performance results, we automatically enumerate all solutions and add in a monitor to collect performance metrics related to size and time. All experiments are performed on Linux Ubuntu 20.04 LTS with
1.8GHz Intel i7 CPU and 16 GB RAM.  
In our experiments, we use a broad mixture of 12 models, including models of data structures (\textbf{arr}, \textbf{btree}, \textbf{dll}, \textbf{sll}), models of java class diagram (\textbf{cd}), models of puzzles  (\textbf{nqueens}), and real world models that were generated automatically from software artifacts (\textbf{CD2Alloy1}, \textbf{CD2Alloy2}, \textbf{diff1}, \textbf{diff2}) and of surgical robots (\textbf{frankenvrep}, \textbf{tombot}).
Table~\ref{tab:size} gives an overview of the size of each model. \textbf{\#Sig} is the number of signatures in the model. As further context, columns \textbf{\#Abs}, \textbf{\#Ext} and \textbf{\#One} show the number of signatures which are abstract, extensions of another signature, and/or use the singleton multiplicity constraint, respectively. Lastly, column \textbf{\#Rel} shows the number of relations declared in the model across all signatures. For each model, we select one existing command to execute and use the scope provided by the published model.

\subsubsection{Baseline}
To help evaluate the impact of \tool's encoding, we compare our CNF encoding to a baseline which uses Alloy's first order logic to explicitly add constraints to the executed command to enforce size. For instance, for the model in Figure~\ref{fig:alloy}, to enumerate scenarios that are of size 2, we can append the following constraints: 

\begin{verbatim}
    (#Node = 2 or #List = 2) and (#Node <= 2) and (#List <= 2)
\end{verbatim}

Our baseline technique automatically generates formulas using the above strategy for all evaluation models.

\subsection{Results}

\subsubsection{RQ1: Overhead}
\tool's encoding focuses on exploring scenarios based on which signature results in the size of the scenario. In contrast, our baseline technique shows the performance for just enumerating scenarios by size. To explore what impact \tool's CNF encoding has, Table~\ref{tab:results-detailed} shows \tool{} and the baseline's performance broken down by each size within scope. Column \textbf{Model} reflects the name of the model under evaluation. Column \textbf{Sz} is the size and column \textbf{\$Scr} is the number of scenarios enumerated at that size. The next five columns relate to the baseline technique's performance. To show the size of the satisfiability problem, columns \textbf{\#PV}, \textbf{\#Var}, \textbf{\#Cls} show the number of primary variables, the number of variables and the number of clauses generated. To show the overhead, columns \textbf{T$_{avg}$} and \textbf{T$_{tot}$} show the average time to discover a scenario and the total time to find all scenarios, respectively. All times are in milliseconds. The next five columns display the same performance information for \tool. If enumerating a size resulted in no scenarios, then that size is not included as a row in the table.

In terms of the size of the satisfiability problem, \tool{} and the baseline often generate similar sized problems, with the baseline generating slightly larger problems with an average of 19 more variables and 92 more clauses, which, on average, makes the baseline problem 1.03$\times$ larger. \tool's smaller sized satisfiability problems highlights how focusing on a specific signature can slightly decrease the complexity of the problem in comparison to the generic baseline encoding that needs to account for broader behavior. In terms of runtime, the two techniques are similar. For 24 and 47 models, \tool{} finishes faster while for 18 models the baseline finishes faster. Moreover, for 40 of the 47 enumerations depicted in Table~\ref{tab:results-detailed}, \tool{} and the baseline technique's total runtime are within 3 seconds of each other. For the remaining seven executions, \tool{} finishes between 3.2 and 18 seconds faster than the baseline technique. These seven executions also highlight how small improvements in the average time to discover a scenario can have an impact when applied across the breadth of all discoverable scenarios. For \textbf{CD2Alloy2}, in which \tool{} finishes 18 seconds faster, \tool{} discovers scenarios on average just 5 milliseconds faster than the baseline. 

While there is not a larger performance difference between the baseline and \tool, the results indicate that \tool's signature-focused encoding can actually help the SAT solver's search performance by having the size of a specific signature be a guiding focus, as indicated by the seven executions in which there was a noticeable difference in runtime performance favoring \tool. Notably, four of these seven executions in which \tool{} has a tangible performance increase over the baseline come from executions over real world models (\textbf{CD2Alloy1}, \textbf{CD2Alloy1}, \textbf{frankenvrep}, \textbf{tombot}). Overall, \tool{} performance in comparison to the baseline indicates that \tool's semi-sorted ordering of signatures is able to both provide a known structure to the discovery of scenarios for the user while also often improving the overall runtime performance.

{
\renewcommand{\arraystretch}{1.1}
\begin{table}[!t]
  \scriptsize
  \centering
  \caption{Comparison of \tool{} and the baseline broken down by each size in scope. All times are in milliseconds (ms). Sizes with zero scenarios are left out.}
  \label{tab:results-detailed}
  \begin{tabular}{l|r|r|rrrrr||rrrrr}
\hline
\multicolumn{1}{c|}{\multirow{2}{*}{\textbf{Model}}} & \multicolumn{1}{c|}{\multirow{2}{*}{\textbf{Sz}}} & \multicolumn{1}{c|}{\multirow{2}{*}{\textbf{\#Scr}}} & \multicolumn{5}{c||}{\textbf{Baseline}} & \multicolumn{5}{c}{\textbf{Reach}} \\ \cline{4-13} 
\multicolumn{1}{c|}{} & \multicolumn{1}{c|}{} & \multicolumn{1}{c|}{} & \multicolumn{1}{c|}{\textbf{\#PV}} & \multicolumn{1}{c|}{\textbf{\#Var}} & \multicolumn{1}{c|}{\textbf{\#Cls}} & \multicolumn{1}{c|}{\textbf{T$_{avg}$}} & \multicolumn{1}{c||}{\textbf{T$_{tot}$}}  & \multicolumn{1}{c|}{\textbf{\#PV}} & \multicolumn{1}{c|}{\textbf{\#Var}} & \multicolumn{1}{c|}{\textbf{\#Cls}} & \multicolumn{1}{c|}{\textbf{T$_{avg}$}} & \multicolumn{1}{c}{\textbf{T$_{tot}$}} \\ \hline
\rowcolor{Gray}
arr & 1 &  255 & 33 & 311 & 663 & 9 & 2543 & 33 & 311 & 664 & 11 & 2839\\ 
\rowcolor{white}
arr & 2 & 1389 & 50 & 502 & 941 & 5 & 7176  & 50 & 498 & 932 & 6 & 8664\\ 
\rowcolor{Gray}
arr & 3 & 2127 & 67 & 674 & 1234 & 4 & 9362  & 67 & 666 & 1213 & 4 & 9515\\ \hline
\rowcolor{white}
btree & 1 & 17 & 20 & 186 & 123 & 1 & 22 & 20 & 186 & 124 & 7 & 121\\ 
\rowcolor{Gray}
btree & 2 & 240 & 44 & 642 & 1614 & 3 & 861 & 44 & 638 & 1605 & 4 & 1089\\ 
\rowcolor{white}
btree & 3 & 560 & 72 & 1362 & 3342 & 4 & 2374 & 72 & 1354 & 3321  & 4 & 2251\\ \hline
\rowcolor{Gray}
cd & 1 & 1 & 1 & 1 & 1 & 0 & 0 & 1 & 1 & 1 & 7 & 7\\ 
\rowcolor{white}
cd & 2 & 1 & 5 & 31 & 37 & 0 & 0 & 5 & 31 & 37 & 7 & 7\\ 
\rowcolor{Gray}
cd & 3 & 2 & 11 & 147 & 237 & 1 & 3 & 11 & 147 & 227 & 6 & 12\\ 
\rowcolor{white}
cd & 4 & 5 & 19 & 318 & 547 & 3 & 15 & 19 & 318 & 541 & 5 & 25\\ 
\rowcolor{Gray}
cd & 5 & 14 & 29 & 712 & 1514 & 5 & 70 & 29 & 712 & 1506 & 4 & 61\\
\rowcolor{white}
cd & 6 & 51 & 41 & 1153 & 2521 & 8 & 424 & 41 & 1153 & 2514 & 5 & 267\\ 
\rowcolor{Gray}
cd & 7 & 190 & 55 & 1736 & 3857 & 10 & 1980 & 55 & 1736 & 3755 & 5 & 1010\\ \hline
\rowcolor{white}
CD2Alloy1 & 5 & 180 & 350 & 6194 & 12851 & 45 & 8224 & 350 & 6170 & 12443 & 22 & 4093\\ 
\rowcolor{Gray}
CD2Alloy1 & 6 & 268 & 480 & 8587 & 18118 & 37 & 10023 & 480 & 8563 & 18016  & 19 & 5318\\ 
\rowcolor{white}
CD2Alloy1 & 7 & 568 & 630 & 11432 & 24363 & 32 & 18362 & 630 & 11420 & 2435 & 22 & 12866\\ 
\rowcolor{Gray}
CD2Alloy1 & 8 & 1171 & 800 & 14845 & 32198 & 31 & 36946 & 800 & 14809 & 32084 & 28 & 33834\\ 
\rowcolor{white}
CD2Alloy1 & 9 & 2433 & 990 & 18640 & 40888 & 41 & 100212 & 990 & 18604 & 40798 & 41 & 100781\\ 
\rowcolor{Gray}
CD2Alloy1 & 1 & 5038 & 1200 & 22791 & 50386 & 56 & 282323 & 1200 & 22755 & 50254 & 56 & 285074\\ \hline
\rowcolor{white}
CD2Alloy2 & 2 & 27 & 36 & 515 & 809 & 1 & 49 & 36 & 508 & 769  & 3 & 96\\ 
\rowcolor{Gray}
CD2Alloy2 & 3 & 21 & 72 & 1128 & 2036 & 5 & 121 & 72 & 1121 & 2032  & 16 & 343\\ 
\rowcolor{white}
CD2Alloy2 & 4 & 177 & 120 & 2004 & 3906 & 4 & 713 & 120 & 1976 & 3819  & 5 & 918\\ 
\rowcolor{Gray}
CD2Alloy2 & 5 & 99 & 180 & 3077 & 6112 & 9 & 924 & 180 & 3049 & 6047  & 9 & 918\\ 
\rowcolor{white}
CD2Alloy2 & 6 & 607 & 252 & 4443 & 9106 & 8 & 4902 & 252 & 4415 & 8993  & 8 & 5300\\ 
\rowcolor{Gray}
CD2Alloy2 & 7 & 285 & 336 & 6006 & 12405 & 10 & 3047 & 336 & 5992 & 12405  & 12 & 3511\\ 
\rowcolor{white}
CD2Alloy2 & 8 & 1538 & 432 & 7819 & 16343 & 16 & 24654 & 432 & 7777 & 16218  & 13 & 20066\\ 
\rowcolor{Gray}
CD2Alloy2 & 9 & 646 & 540 & 9857 & 20655 & 22 & 14253 & 540 & 9815 & 20559  & 19 & 12458\\ 
\rowcolor{white}
CD2Alloy2 & 10 & 3255 & 660 & 12114 & 25520 & 25 & 83315 & 660 & 12072 & 25376  & 20 & 65241\\ \hline
\rowcolor{Gray}
diff1 & 5 & 150 & 355 & 7262 & 16299 & 24 & 3662 & 355 & 7234 & 15823 & 150  & 2480\\ 
\rowcolor{white}
diff1 & 6 & 235 & 486 & 10022 & 22888 & 31 & 7413 & 486 & 9994 & 22769 & 235  & 4897\\ \hline
\rowcolor{Gray}
diff2 & 5 & 24 & 355 & 7262 & 16277 & 20 & 493 & 355 & 7234 & 15801 & 24  & 967\\ 
\rowcolor{white}
diff2 & 6 & 46 & 486 & 10022 & 22850 & 35 & 1625 & 486 & 9994 & 22731  & 32 & 1508\\ \hline
\rowcolor{Gray}
dll & 1 & 17 & 20 & 332 & 1052 & 2 & 43 & 20 & 332 & 1053  & 4 & 71\\ 
\rowcolor{white}
dll & 2 & 120 & 44 & 897 & 2612 & 5 & 636 & 44 & 893 & 2603  & 3 & 455\\ 
\rowcolor{Gray}
dll & 3 & 560 & 72 & 1588 & 4554 & 6 & 3711 & 72 & 1580 & 4533 & 4 & 2769\\ \hline
\rowcolor{white}
frankenvrep & 4 & 1200 & 134 & 1372 & 2820 & 7 & 8727 & 134 & 1232 & 1200 & 4 & 5804\\ 
\rowcolor{Gray}
frankenvrep & 5 & 2000 & 163 & 1785 & 3847 & 7 & 14323 & 172 & 1749 & 3630  & 5 & 10654\\ \hline
\rowcolor{white}
nqueens & 1 & 0 & 0 & 0 & 0 & 0 & 0 & 0 & 0 & 0 & 0 & 0\\ 
\rowcolor{Gray}
nqueens & 1 & 1 & 33 & 364 & 1097 & 8 & 8 & 33 & 364 & 1097 & 23 & 23\\ 
\rowcolor{white}
nqueens & 4 & 2 & 132 & 2864 & 8042 & 76 & 152  & 132 & 2859 & 8036 & 95 & 190\\ 
\rowcolor{Gray}
nqueens & 5 & 10 & 165 & 4061 & 11365 & 57 & 576 & 165 & 4057 & 11362 & 51 & 517\\ \hline
\rowcolor{white}
sll & 0 & 1 & 0 & 0 & 0 & 0 & 0 & 0 & 0 & 0 & 0 & 0\\ 
\rowcolor{Gray}
sll & 1 & 6 & 4 & 14 & 15 & 1 & 6 & 4 & 13 & 14 & 2 & 17\\ 
\rowcolor{white}
sll & 2 & 38 & 12 & 124 & 176 & 1 & 71 & 12 & 115 & 156 & 2 & 78\\ 
\rowcolor{Gray}
sll & 3 & 299 & 24 & 339 & 527 & 2 & 655 & 24 & 322 & 482 & 2 & 607\\ \hline
\rowcolor{white}
tombot & 4 & 3000 & 134 & 1372 & 2822 & 6 & 19156 & 134 & 1232 & 2318 & 4 & 14804\\ 
\rowcolor{Gray}
tombot & 5 & 2760 & 163 & 1785 & 3847 & 5 & 16384 & 172 & 1749 & 3630  & 5 & 14513\\ \hline
\end{tabular}
    \vspace{3ex}
\end{table}
}

\subsubsection{RQ2: Comparison to the Analyzer's Enumeration}
\tool's enumeration strategy is a tradeoff. We append additional CNF clauses that must be satisfied by any scenario enumerated, which increases the complexity of the satisfiability problem. However, in exchange, we are able to tailor the generated satisfiability problem to a specific size. 
To see the impact, Table~\ref{tab:comparison} shows a comparison between the default Analyzer, the baseline technique and \tool. Column \textbf{Model} reflects the name of the model under evaluation. Column \textbf{Scp} is the size and column \textbf{\$Scr} is the number of scenarios enumerated at that size. The next five columns relate to the Analyzer's performance. To show the size of the satisfiability problem, columns \textbf{\#PV}, \textbf{\#Var}, \textbf{\#Cls} show the number of primary variables, the number of variables and the number of clauses generated. To show the overhead, columns \textbf{T$_{avg}$} and \textbf{T$_{tot}$} show the average time to discover a scenario and the total time to find all scenarios, respectively. For the baseline and \tool, we present summary information. Under the baseline and \tool's headers, the column \textbf{T$_{avg}$} is the average time to discover a scenario across all scenarios of all sizes and column \textbf{T$_{tot}$} reflects the total runtime summed across all sizes.

{
\renewcommand{\arraystretch}{1.1}
\begin{table}[!t]
  \scriptsize
  \centering
  \caption{Comparison of performance across the entire scope. All times are in ms.}
  \label{tab:comparison}
  \begin{tabular}{l|r|r|rrrrr||rr||rr}
\hline
\multicolumn{1}{c|}{\multirow{2}{*}{\textbf{Model}}} & \multicolumn{1}{c|}{\multirow{2}{*}{\textbf{Scp}}} & \multicolumn{1}{c|}{\multirow{2}{*}{\textbf{\#Scr}}} & \multicolumn{5}{c||}{\textbf{Alloy}} & \multicolumn{2}{c||}{\textbf{Baseline}} & \multicolumn{2}{c}{\textbf{Reach}} \\ \cline{4-12} 
\multicolumn{1}{c|}{} & \multicolumn{1}{c|}{} & \multicolumn{1}{c|}{} & \multicolumn{1}{c|}{\textbf{\#PV}} & \multicolumn{1}{c|}{\textbf{\#Var}} & \multicolumn{1}{c|}{\textbf{\#Cls}} & \multicolumn{1}{c|}{\textbf{T$_{avg}$}} & \multicolumn{1}{c||}{\textbf{T$_{tot}$}} & \multicolumn{1}{c|}{\textbf{T$_{avg}$}} & \multicolumn{1}{c||}{\textbf{T$_{tot}$}} & \multicolumn{1}{c|}{\textbf{T$_{avg}$}} & \multicolumn{1}{c}{\textbf{T$_{tot}$}}\\ \hline
\rowcolor{Gray}
arr & 3 & 3771 & 67 & 666 & 1210 & 4 & 18529 & 5 & 19081 & 5 & 21018\\ 
\rowcolor{white}
btree & 3 & 817  & 72 & 1354 & 3318 & 4 & 4042 & 3 & 3257 & 4 & 3461\\ 
\rowcolor{Gray}
cd & 7 & 264 & 55 & 1706 & 3749 & 6 & 1834 & 9 & 2492 & 5 & 1389\\ 
\rowcolor{white}
CD2Alloy1 & 10 & 9658 & 1200 & 22438 & 48988 & 70 & 678509 & 47 & 456090 & 45 & 441966\\ 
\rowcolor{Gray}
CD2Alloy2 & 10 & 6655 & 660 & 11702 & 23889 & 26 & 176923 & 19 & 131978 & 16 & 108851\\ 
\rowcolor{white}
diff1 & 6 & 385 & 486 & 9813 & 22097 & 21 & 8178 & 28 & 11075 & 19 & 7377\\ 
\rowcolor{Gray}
diff2 & 6 & 70 & 486 & 9813 & 22059  & 33 & 2357 & 30 & 2118 & 35 & 2475\\ 
\rowcolor{white}
dll & 3 & 697 & 72 & 1580 & 4530 & 4 & 3083 & 6 & 4390 & 4 & 3295\\ 
\rowcolor{Gray}
frankenvrep & 5 & 3200 & 163 & 1593 & 3191 & 4 & 14802 & 7 & 23050 & 5 & 16458\\ 
\rowcolor{white}
nqueens & 5 & 14 & 165 & 4057 & 11357 & 113 & 1595 & 52 & 736 & 52 & 730\\ 
\rowcolor{Gray}
sll & 3 & 344 & 24 & 325 & 485 & 1 & 581 & 2 & 732 & 2 & 702\\ 
\rowcolor{white}
tombot & 5 & 5760 & 163 & 1593 & 3191 & 5 & 28809 & 6 & 35540 & 5 & 29317\\ \hline
\end{tabular}
  \vspace{3ex}
\end{table}
}

The total number of scenarios is constant across all three techniques, as all information is related to enumerating all scenarios from size 0 to the upper bound. For 10 of the 12 models, the total runtime of all three techniques is within 3 seconds of each other. Therefore, we do not suffer any negative performance overhead when enumerating scenarios by size rather than allowing the SAT solver to determine the order. The most noticeable difference in performance appears for model \textbf{CD2Alloy1}, which is an automatically generated model from a UML class diagram. For \textbf{CD2Alloy1}, both the baseline technique and \tool{} reduce the average time to discover a scenario, which has a ripple effect leading to the techniques finishing nearly 4 minutes faster than the default Analyzer. Similarly, for \textbf{CD2Alloy2}, both techniques finish about a minute faster the Analyzer. Of note, for the \textbf{nqueens} model, the baseline and \tool{} both noticeably reduce the average time to discover a scenario; however, the difference in overall runtime compared to the Analyzer is less than a second due to the small number of discoverable scenarios. 
\section{Case Study - Iterative Deepening}
Within Alloy, it is standard practice to start with a small scope, and then to increase the scope and re-run the Analyzer to build confidence. This iterative deepening process produces largely redundant work, as the scenarios found for smaller, previously explored scopes must be re-discovered when the scope increases. To illustrate, when a scope $s$ is increased to $s + 1$, the Analyzer regenerates all the scenarios related to scopes of size 0, 1, 2, $\ldots$, $s$ and additionally generates scenarios for scope $s + 1$. However, only the new scenarios of size $s + 1$ are actually \emph{new} for the user.

A study of how user interact with the Analyzer reveals that users often perform consecutive executions with models that only vary slightly: 59.9\% only differ by one formula, 17.6\% only differ in the scope, and 11.8\% are redundant to past executions~\cite{alloyuses}. Therefore, \tool's ability to enumerate scenarios by size, if combined with a monitor to detect if no change or only a scope level change, can improve the Analyzer's analysis for 26.4\% of Alloy executions. To investigate how effective \tool{} can be at providing native support for iterative deepening, we simulate the process by using \tool's infrastructure to enumerate scenarios for sizes $s + 1$ and $s + 2$ for our evaluation models from Section~\ref{sec:eval}.

\subsection{Results}

Table~\ref{tab:itrdeep} shows the comparison between the Analyzer's enumeration, the baseline technique and \tool's performance in an iterative deepening setting. Column \textbf{Model} reflects the name of the model under evaluation and column \textbf{Scp} is the scope. The next 18 columns display performance information for the different techniques. Specifically, performance information for each technique is given in 6 columns with the following meanings: columns \textbf{\#PV}, \textbf{\#Var}, \textbf{\#Cls} show the number of primary variables, the number of variables and the number of clauses generated; column \textbf{\#Scr} is the number of scenarios enumerated; and to show the overhead, columns \textbf{T$_{avg}$} and \textbf{T$_{tot}$} show the average time to discover a scenario and the total time to find all scenarios, respectively. All times are in milliseconds. 

\begin{sidewaystable}[!hp]
\renewcommand{\arraystretch}{1.1}
  \scriptsize
  \centering
  \caption{Iterative deepening results for the Analyzer, the baseline technique and \tool. All times are in ms.}
  \label{tab:itrdeep}
  \begin{tabular}{l|c|rrrrrr|rrrrrr|rrrrrr}
\hline
\multicolumn{1}{c|}{\multirow{2}{*}{\textbf{Model}}} & \multicolumn{1}{c|}{\multirow{2}{*}{\textbf{Scp}}} & \multicolumn{6}{c|}{\textbf{Alloy}}                                                                                                                                                                                                    & \multicolumn{6}{c|}{\textbf{Baseline}}                                                                                                                                                                                                 & \multicolumn{6}{c}{\textbf{Reach}}                                                                                                                                                                                                    \\ \cline{3-20} 
\multicolumn{1}{c|}{}                                & \multicolumn{1}{c|}{}                                & \multicolumn{1}{c|}{\textbf{\#PV}} & \multicolumn{1}{c|}{\textbf{\#Var}} & \multicolumn{1}{c|}{\textbf{\#Cls}} & \multicolumn{1}{c|}{\textbf{\#Scn}} & \multicolumn{1}{c|}{\textbf{T$_avg$}} & \multicolumn{1}{c|}{\textbf{T$_tot$}} & \multicolumn{1}{c|}{\textbf{\#PV}} & \multicolumn{1}{c|}{\textbf{\#Var}} & \multicolumn{1}{c|}{\textbf{\#Cls}} & \multicolumn{1}{c|}{\textbf{\#Scn}} & \multicolumn{1}{c|}{\textbf{T$_avg$}} & \multicolumn{1}{c|}{\textbf{T$_tot$}} & \multicolumn{1}{c|}{\textbf{\#PV}} & \multicolumn{1}{c|}{\textbf{\#Var}} & \multicolumn{1}{c|}{\textbf{\#Cls}} & \multicolumn{1}{c|}{\textbf{\#Scn}} & \multicolumn{1}{c|}{\textbf{T$_avg$}} & \multicolumn{1}{c}{\textbf{T$_tot$}} \\ \hline
\rowcolor{Gray}
\rowcolor{Gray}
arr & 4 & 84 & 817 & 1473 & 4977 & 5 & 28314 & 84 & 833 & 1528 & 1206 & 6 & 7458 & 84 & 829 & 1519 & 1206 & 6 & 7810\\
\rowcolor{white}
arr & 5 & 101 & 968 & 1736 & 5265 & 7 & 41602 & 101 & 989 & 1813 & 288 & 4 & 1355 & 101 & 986 & 1810 & 288 & 3 & 1144\\ \hline
\rowcolor{Gray}
btree & 4 & 104 & 2269 & 5516 & 8097 & 6 & 48858 & 104 & 2285 & 5571 & 7280 & 5 & 38565 & 104 & 2281 & 5562 & 7280 & 5 & 38735\\ 
\rowcolor{white}
btree & 5 & 140 & 3708 & 9174 & 34305 & 9 & 310767 & 140 & 3729 & 9251 & 26208 & 8 & 212916 & 140 & 3726 & 9248 & 26208 & 7 & 200271\\ \hline
\rowcolor{Gray}
cd & 8 & 71 & 2440 & 5453 & 1110 & 8 & 9841 & 71 & 2479 & 5603 & 846 & 6 & 5191 & 71 & 2479 & 5612 & 846 & 7 & 6470\\ 
\rowcolor{white}
cd & 9 & 89 & 4197 & 10735 & 4905 & 12 & 60600 & 89 & 4238 & 10894 & 3795 & 9 & 34784 & 89 & 4238 & 10904 & 3795 & 9 & 35862\\ \hline
\rowcolor{Gray}
CD2Alloy1 & 11 & 1430 & 26861 & 58798 & 19668 & 89 & 1755121 & 1430 & 26988 & 59302 & 10010 & 73 & 733423 & 1430 & 27220 & 60244 & 10010 & 71 & 719812\\ 
\rowcolor{white}
CD2Alloy1 & 12 & 1680 & 31610 & 69178 & 39377 & 105 & 4158553 & 1680 & 31755 & 69764 & 19709 & 87 & 1721733 & 1680 & 32011 & 70804 & 19709 & 86 & 1704248\\ \hline
\rowcolor{Gray}
CD2Alloy2 & 11 & 792 & 14157 & 29069 & 7920 & 28 & 227959 & 792 & 14348 & 29825 & 1265 & 27 & 35356 & 792 & 14576 & 30767 & 1265 & 33 & 42644\\
\rowcolor{white}
CD2Alloy2 & 12 & 936 & 16990 & 35261 & 14027 & 43 & 607030 & 936 & 17208 & 36140 & 6107 & 37 & 225998 & 936 & 17458 & 37170 & 6107 & 37 & 227464\\ \hline
\rowcolor{Gray}
diff1 & 7 & 637 & 13040 & 29732 & 944 & 24 & 23303 & 637 & 13105 & 29968 & 559 & 27 & 15301 & 637 & 13256 & 30551 & 559 & 28 & 16111\\ 
\rowcolor{white}
diff1 & 8 & 808 & 16849 & 38935 & 2061 & 37 & 78183 & 808 & 16932 & 39251 & 1117 & 38 & 42895 & 808 & 17100 & 39908 & 1117 & 40 & 45454\\ \hline
\rowcolor{Gray}
diff2 & 7 & 637 & 13040 & 29674 & 80 & 45 & 3641 & 637 & 13105 & 29910 & 10 & 56 & 565 & 637 & 13256 & 30493 & 10 & 169 & 1697\\ 
\rowcolor{white}
diff2 & 8 & 808 & 16849 & 38853 & 180 & 65 & 11809 & 808 & 16932 & 39169 & 100 & 51 & 5102 & 808 & 17100 & 39826 & 100 & 60 & 6036\\ \hline
\rowcolor{Gray}
dll & 4 & 104 & 2398 & 6831 & 2517 & 8 & 20475 & 104 & 2414 & 6886 & 1820 & 5 & 10106 & 104 & 2410 & 6877 & 1820 & 5 & 10251\\ 
\rowcolor{white}
dll & 5 & 140 & 3532 & 10132 & 6885 & 9 & 63176 & 140 & 3553 & 10209 & 4368 & 6 & 29108 & 140 & 3550 & 10206 & 4368 & 7 & 31613\\ \hline
\rowcolor{Gray}
frankenvrep & 6 & 196 & 2063 & 4234 & 5760 & 6 & 39462  & 196 & 2154 & 4593 & 2560 & 5 & 15031 & 214 & 2326 & 4945 & 2560 & 5 & 14909\\ 
\rowcolor{white}
frankenvrep & 7 & 233 & 2608 & 5452 & 8960 & 7 & 67542 & 233 & 2704 & 5784 & 3200 & 5 & 18958 & 260 & 2968 & 6355 & 3200 & 6 & 19925\\ \hline
\rowcolor{Gray}
nqueens & 6 & 198 & 5403 & 15088 & 18 & 142 & 2572 & 198 & 5407 & 15097 & 4 & 135 & 540 & 198 & 5404 & 15099 & 4 & 192 & 770\\ 
\rowcolor{white}
nqueens & 7 & 231 & 6918 & 19277 & 58 & 117 & 6810 & 231 & 6921 & 19282 & 40 & 57 & 2312 & 231 & 6919 & 19289 & 40 & 63 & 2536\\ \hline
\rowcolor{Gray}
sll & 4 & 40 & 626 & 981 & 3425 & 3 & 12694 & 40 & 659 & 1091 & 3081 & 2 & 7699 & 40 & 651 & 1069 & 3081 & 3 & 10590\\ 
\rowcolor{white}
sll & 5 & 60 & 1215 & 1992 & 45624 & 7 & 343614 & 60 & 1258 & 2146 & 42199 & 7 & 325469 & 60 & 1252 & 2135 & 42199 & 5 & 248271\\ \hline
\rowcolor{Gray}
tombot & 6 & 196 & 2063 & 4234 & 8960 & 6 & 55663 & 196 & 2154 & 4593 & 3200 & 7 & 22788 & 214 & 2326 & 4945 & 3200 & 6 & 22000\\ 
\rowcolor{white}
tombot & 7 & 233 & 2608 & 5452 & 12160 & 7 & 85414 & 233 & 2704 & 5784 & 3200 & 7 & 24925 & 260 & 2968 & 6355 & 3200 & 7 & 23571\\ \hline
\end{tabular}
\end{sidewaystable}

Since the Analyzer can only treat the scope of a command as an upper bound, for each model, the Analyzer discovers on average 3.0$\times$ more scenarios and takes on average 3.6$\times$ longer than both the baseline and \tool{}. The Analyzer does create the smallest satisfiability problem, although both the baseline and \tool{} only create on average 1.02$\times$ and 1.05$\times$  larger models. Across most models, the baseline technique and \tool{} have similar performance. By design, both techniques generate the same number of scenarios. For 19 of the 24 executions, \tool{} and baseline finish within 3 seconds of each other. However, there are two notable points of difference. First, for models with a very large number of scenarios, \tool{} outperforms the baseline. To highlight, for the data structure models (\textbf{btree}, \textbf{sll}), \tool{} finishes 12.6 and 77.2 seconds faster. For both of these models, the constraints allow for a very broad range of behavior, as indicated by the extremely large number of scenarios discovered (26208 and 42199, respectively). Additionally, for both iterative deepening explorations of the \textbf{CD2Alloy1} model, \tool{} finishes 13 and 17 seconds faster than the baseline. The first iterative deepening generates just over 10,000 scenarios and the second generates just under 20,000 scenarios. Second, for one model, \textbf{CD2Alloy2}, the baseline finds all 1265 scenarios of size 11 about 7 seconds faster than \tool. 

Overall, the results re-iterate that \tool's encoding provides an efficient enumeration despite adding constraints to enforce size through targeting specific signatures. At the same time, the results indicate that as the space of valid scenarios increases, \tool{} enforcement of a soft ordering on signatures can have a positive impact on the overall enumeration runtime. As a result, \tool's enumeration strategy is a good candidate to support iterative deepening. 

\section{Related Work}
Our technique is closely related to techniques which looks to enhance the Analyzer's scenario enumeration process. One traditional approach is to reduce the number of scenarios through symmetry  breaking, where constraints are added to the formula to remove isomorphic solutions~\cite{Crawford92,Shlyakhter01EffectiveSymmetryBreaking,KhurshidETALSAT2003}. 
Beyond symmetry breaking, several past projects improve scenario enumeration by trying to narrow what scenarios are generated using a specific criteria, e.g., abstract functions~\cite{Seabs}, minimality~\cite{Alluminum}, field exhaustiveness~\cite{PonzioETALFSE2016}, and coverage~\cite{AUnit,PorncharoenwaseETALCompoSat2018}. All of these techniques reduce the number of scenarios by applying some criteria across the entire enumeration. 
However, the order of scenarios generated is still dictated by the SAT solver's order of discovery. 
Another approach to enhancing scenario enumeration is to focus on the order in which scenarios are presented. Hawkeye is a tool which allows for users to give on-the-fly guidance to direct the scenario enumeration process by specifying how the next scenario should differ with respect to the current scenario~\cite{Hawkeye}.  Our approach can work in tandem with these enumeration strategies to take advantage of their reduction in the scenario search space while \tool{} provides a semi-sorted order and efficient staged generation.

Our case study of \tool{} targets improving the performance of the Analyzer in an iterative deepening setting, which is an incremental analysis problem. 
There are three main bodies of work related to incremental Alloy analysis. First, Titanium uses all the scenarios of the previous model version to calculate tighter bounds for relational variables for the next iteration~\cite{titanium}. 
Second, iAlloy uses static analysis to determine which commands to avoid re-executing and determine which scenarios to reuse, with a focus on formula level changes~\cite{iAlloy}. 
Third, Platinum slices Alloy models at a boolean level using the CNF formula and reuses scenarios if a redundant CNF slice is detected, and focuses on model structure changes~\cite{Platinum}. All three efforts are intended to be more generic than \tool, working with a broader range of incremental changes. In contrast, \tool{} is specialized to focus on streamlining the Analyzer's incremental performance related to scope-level changes. However, since our technique is built into the CNF encoding used to execute commands, we believe \tool{} can be integrated with these techniques to help improve their incremental analysis.

In general, beyond Alloy, researchers have focused on improving scenario enumeration strategies, .e.g. dedicated
search~\cite{BoyapatiETALISSTA2002}, mixing of generators and
solvers \cite{UDITA,KurajETALOOPSLA2013}, solver-aided
languages~\cite{RingerETALOOPSLA2017}, and
sampling~\cite{MeelETAL2016,DutraETALSMTSampler2018}. We believe that improvements to the scenarios that get generated by the Analyzer could be refined by some of these approaches, and further combined with \tool. Additionally, researchers have developed novel verification efforts which utilize scenario enumerating toolsets, e.g. automated test input generation~\cite{MarinovKhurshid01TestEra}, and model counting for reliability analysis~\cite{FilieriETALICSE2013}.  In future work, we plan to explore how \tool's enumeration strategy can benefit the broader adoption of these automated verification efforts, in particular in an incremental setting by allowing the user to discover scenarios of specific sizes for testing. 
\section{Conclusion and Future Work}
A key strength of Alloy is its scenario enumeration toolset the Analyzer. Unfortunately, for a scenario finding toolset to be valuable, the user needs to be empowered to efficiently explore scenarios. The Analyzer's enumeration process is only guided by a back-end SAT solver. To address this limitation, this paper introduces \tool, a novel framework for enumerating Alloy scenarios by size. To achieve this, \tool{} utilizes a CNF encoding that prioritizes the size of specific signatures, which helps to further order the scenarios. Our experimental results highlight that our modified CNF encoding does not create a burdensome overhead and, for models with a large scenario space, \tool's encoding noticeably improves the discovery time. Furthermore, our case study explores how \tool's functionality can be applied to efficiently provide an execution environment that supports iterative deepening, which is a common model development process in Alloy. In the future, we plan to combine \tool{} with our command dependency graph from our Alloy repair work in order to be able to automatically detect when a given execution is performing iterative deepening and to the allow users to specify their own ordering to the signatures.

%
%
%
\bibliographystyle{splncs04}
\bibliography{bib}

\begin{thebibliography}{10}
\providecommand{\url}[1]{\texttt{#1}}
\providecommand{\urlprefix}{URL }
\providecommand{\doi}[1]{https://doi.org/#1}

\bibitem{AlloyWebsite}
{Alloy analyzer Website}: \url{http://alloytools.org} (2019)

\bibitem{titanium}
Bagheri, H., Malek, S.: Titanium: efficient analysis of evolving alloy
  specifications. In: Proceedings of the 2016 24th ACM SIGSOFT Symposium on the
  Foundations of Software Engineering (FSE). pp. 27--38 (2016)

\bibitem{BoyapatiETALISSTA2002}
Boyapati, C., Khurshid, S., Marinov, D.: {K}orat: Automated testing based on
  {J}ava predicates. In: ISSTA (2002)

\bibitem{beafix}
Brida, S.G., Regis, G., Zheng, G., Bagheri, H., Nguyen, T., Aguirre, N., Frias,
  M.F.: Bounded exhaustive search of alloy specification repairs. In: 43rd
  {IEEE/ACM} International Conference on Software Engineering, {ICSE} 2021,
  Madrid, Spain, 22-30 May 2021. pp. 1135--1147 (2021)

\bibitem{CheckMateGitHub}
{CheckMate GitHub}: \url{https://github.com/ctrippel/checkmate} (2019)

\bibitem{Crawford92}
Crawford, J.: A theoretical analysis of reasoning by symmetry in first-order
  logic (extended abstract). In: AAAI-92 Workshop on Tractable Reasoning (1992)

\bibitem{DutraETALSMTSampler2018}
Dutra, R., Bachrach, J., Sen, K.: {SMTSampler}: Efficient stimulus generation
  from complex {SMT} constraints. In: {ICCAD} (2018)

\bibitem{FilieriETALICSE2013}
Filieri, A., Pasareanu, C.S., Visser, W.: Reliability analysis in {Symbolic}
  {PathFinder}. In: {ICSE} (2013)

\bibitem{DynAlloy}
Frias, M.F., Galeotti, J.P., Pombo, C.G.L., Aguirre, N.M.: {DynAlloy}:
  Upgrading {Alloy} with actions. In: ICSE (2005)

\bibitem{UDITA}
Gligoric, M., Gvero, T., Jagannath, V., Khurshid, S., Kuncak, V., Marinov, D.:
  Test generation through programming in {UDITA}. In: ICSE (2010)

\bibitem{JacksonAlloy2002}
Jackson, D.: {Alloy}: A lightweight object modelling notation. IEEE
  Transactions on Software Engineering (TSE)  \textbf{11},  256--290 (2002)

\bibitem{JacksonAlloyBook2006}
Jackson, D.: Software Abstractions: Logic, Language, and Analysis. The MIT
  Press (2006)

\bibitem{KhurshidETALSAT2003}
Khurshid, S., Marinov, D., Shlyakhter, I., Jackson, D.: A case for efficient
  solution enumeration. In: SAT (2003)

\bibitem{KurajETALOOPSLA2013}
Kuraj, I., Kuncak, V., Jackson, D.: Programming with enumerable sets of
  structures. In: {OOPSLA} (2015)

\bibitem{alloyuses}
Li, X., Shannon, D., Walker, J., Khurshid, S., Marinov, D.: Analyzing the uses
  of a software modeling tool. Electron. Notes Theor. Comput. Sci.
  \textbf{164}(2),  3--18 (2006)

\bibitem{CD2Alloy}
Maoz, S., Ringert, J.O., Rumpe, B.: {CD2Alloy: C}lass diagrams analysis using
  {Alloy} revisited. In: MODELS (2011)

\bibitem{CDDiff}
Maoz, S., Ringert, J.O., Rumpe, B.: {CDDiff: S}emantic differencing for class
  diagrams. In: ECOOP (2011)

\bibitem{MarinovKhurshid01TestEra}
Marinov, D., Khurshid, S.: {TestEra}: A novel framework for automated testing
  of {Java} programs. In: ASE (2001)

\bibitem{MeelETAL2016}
Meel, K.S., Vardi, M.Y., Chakraborty, S., Fremont, D.J., Seshia, S.A., Fried,
  D., Ivrii, A., Malik, S.: Constrained sampling and counting: Universal
  hashing meets {SAT} solving. In: Beyond NP, Papers from the 2016 {AAAI}
  Workshop (2016)

\bibitem{Amalgam}
Nelson, T., Danas, N., Dougherty, D.J., Krishnamurthi, S.: The power of "why"
  and "why not": Enriching scenario exploration with provenance. In: FSE (2017)

\bibitem{Alluminum}
Nelson, T., Saghafi, S., Dougherty, D.J., Fisler, K., Krishnamurthi, S.:
  Aluminum: Principled scenario exploration through minimality. In: ICSE (2013)

\bibitem{Margrave}
Nelson, T., Barratt, C., Dougherty, D.J., Fisler, K., Krishnamurthi, S.: The
  {Margrave} tool for firewall analysis. In: LISA (2010)

\bibitem{PonzioETALFSE2016}
Ponzio, P., Aguirre, N., Frias, M.F., Visser, W.: Field-exhaustive testing. In:
  {FSE} (2016)

\bibitem{PorncharoenwaseETALCompoSat2018}
Porncharoenwase, S., Nelson, T., Krishnamurthi, S.: {CompoSAT}:
  Specification-guided coverage for model finding. In: {FM} (2018)

\bibitem{RingerETALOOPSLA2017}
Ringer, T., Grossman, D., Schwartz-Narbonne, D., Tasiran, S.: A solver-aided
  language for test input generation. OOPLSA  (2017)

\bibitem{OpenflowAlloy}
Ruchansky, N., Proserpio, D.: A (not) {NICE} way to verify the {Openflow}
  switch specification: Formal modelling of the {Openflow} switch using
  {Alloy}. SIGCOMM  (2013)

\bibitem{Shlyakhter01EffectiveSymmetryBreaking}
Shlyakhter, I.: Generating effective symmetry-breaking predicates for search
  problems. In: SAT (2001)

\bibitem{Hawkeye}
Sullivan, A.: Hawkeye: User guided enumeration of scenarios. In: ISSRE (2021)

\bibitem{Seabs}
Sullivan, A., Marinov, D., Khurshid, S.: Solution enumeration abstraction - {A}
  modeling idiom to enhance a lightweight formal method. In: ICFEM (2019)

\bibitem{Sullivan2017EvaluatingSM}
Sullivan, A., Wang, K., Khurshid, S., Marinov, D.: Evaluating state modeling
  techniques in alloy. In: SQAMIA (2017),
  \url{http://ceur-ws.org/Vol-1938/paper-sul.pdf}

\bibitem{AUnit}
Sullivan, A., Wang, K., Zaeem, R.N., Khurshid, S.: Automated test generation
  and mutation testing for {A}lloy. In: ICST (2017)

\bibitem{KodKod}
Torlak, E., Jackson, D.: Kodkod: A relational model finder. In: TACAS (2007)

\bibitem{CheckMateMICRO2018}
Trippel, C., Lustig, D., Martonosi, M.: {CheckMate}: Automated synthesis of
  hardware exploits and security litmus tests. In: MICRO (2018)

\bibitem{CheckMateMicro2019}
Trippel, C., Lustig, D., Martonosi, M.: Security verification via automatic
  hardware-aware exploit synthesis: The {CheckMate} approach. {IEEE} Micro
  (2019)

\bibitem{ARepair}
Wang, K., Sullivan, A., Khurshid, S.: Automated model repair for {A}lloy. In:
  ASE (2018). \doi{10.1145/3238147.3238162}

\bibitem{iAlloy}
Wang, W., Wang, K., Gligoric, M., Khurshid, S.: Incremental analysis of
  evolving alloy models. In: Vojnar, T., Zhang, L. (eds.) Tools and Algorithms
  for the Construction and Analysis of Systems. pp. 174--191 (2019)

\bibitem{Platinum}
Zheng, G., Bagheri, H., Rothermel, G., Wang, J.: Platinum: Reusing constraint
  solutions in bounded analysis of relational logic. In: Fundamental Approaches
  to Software Engineering. pp. 29--52 (2020)

\end{thebibliography}

\end{document}